\documentclass[11pt,twoside]{article} \usepackage{fleqn,espcrc1}

\newcommand{\AmS}{{\protect\the\textfont2
  A\kern-.1667em\lower.5ex\hbox{M}\kern-.125emS}}

\title{Virtual Color Superconductivity and Nucleon Structure\footnote
{Talk given at the {\em International Conference on Quark Nuclear Physics},
 Adelaide, Australia, Feb. 21-25, 2000.}}

\author{S. Ying\address{Research Center for Theoretical Physics, 
        Department of Physics, Fudan University\\
        Shanghai 200433, China}}
       
\begin{document}

\maketitle

\begin{abstract}
   The observed electromagnetic (EM) properties of a nucleon at high
energy are studied to search for a possible metastable color
superconducting phase, called a virtual phase, that has an energy
density close to the true ground state of the vacuum. The discussion
is based on a mechanism of spontaneous partial breaking of the EM
$U(1)$ gauge symmetry inside a color superconductor. Favorable
evidences for such a scenario are found.
\end{abstract}

\section{Introduction}

   Theoretical and empirical studies seem to favor \cite{review} the
possibility that there exists a metastable color superconducting phase
with its energy density close to the chiral symmetry breaking phase of
the strong interaction vacuum. Since the color superconducting phase
breaks the EM $U(1)$ gauge symmetry spontaneously, it is quite likely
that it could be found from the response of a nucleon to external EM
probes. The questions are 1) why it is detectable 2) in which way it
can be seen and 3) does it correspond to reality?  In this talk, I
would like to introduce a recent work done on this subject
\cite{DISpap}.

  The problems concerning the EM properties of a nucleon that this
work try to provide a logically coherent picture are the following
\begin{enumerate}
\item Possible violation of the GDH sum rule ($\sim 10\%$), which is
      still not saturated by the data at the present \cite{gdht}.
\item The rapid rise of the structure functions $F_2(x)$ and $g_1(x)$
      for a nucleon at small x measured at HERA and SMC. It seems to
      lead to a violation of Froissart bound.
\item The difference between $F_2(x)$ extracted from charged lepton DIS
      and the one from the neutrino DIS.
\item And others \cite{DISpap}.
\end{enumerate}
  Due to the length limit, I can only briefly discuss some of them and
suppress most of the known references except those new ones. They can
be found in Refs.  \cite{review,DISpap}. The basic assumptions
sufficient for the present talk are 1) there is a virtual color
superconducting phase for the strong interaction vacuum and 2)
together with the normal quasi-particles of the true vacuum phase, the
quasi-particles of the virtual phase has a small probability to
participate in the high energy processes.  They can be substantiated
by more solid theoretical argumentation \cite{review}, but we have no
time for that here. Therefore a nucleon and any hadrons containing up
and down quarks can become superconducting with a small probability
\cite{review}.  This superconducting aspect of the nucleon is studied
in more detail in the following. Our emphasis is on those properties
of the color superconducting phase that are absent in the normal
chiral symmetry breaking phase.

\section{Partial Breaking of EM Gauge Symmetry}
 
 In a spontaneous breaking of a global symmetry, massless Goldstone
bosons appear following the Goldstone theorem. What is less mentioned
in literature is the separation of the corresponding charge that
generates the symmetry. In a
color superconductor, the EM vertex can be written as
\begin{equation}
     J_{em}^\mu = J_{core}^\mu + {l^\mu\over q\cdot l}J_{spread},
\label{test1}
\end{equation}
where $J_{core}^\mu$ is the contribution from the ``core'' part of the
charge on the particle and the scalar function $J_{spread}$ is the
strength of the spreaded component of the charge that is carried away
by the massless Goldstone boson. $J_{spread}$ is related to the mass
matrix that breaks the symmetry via a Ward identity. The 4-vector
$l^\mu$ is the longitudinal polarization of the Goldstone boson in a
momentum carrying ``medium'', like a nucleon, ($l^\mu=q^\mu$ in the
vacuum). Although $q_\mu J_{em}^\mu=0$ even in the symmetry breaking
phase due to gauge invariance, $J^\mu_{core}$, which contribute to the
observables in DIS \cite{DISpap}, satisfies
\begin{equation}
    q_\mu J^\mu_{core} = -J_{spread} \ne 0.
\label{test2}
\end{equation}

If the symmetry is local, like the EM $U(1)$ gauge symmetry, then
Higgs mechanism prevents 
the Goldstone boson to participate in physical collisions, which is well 
stated in text books. The effects of the Goldstone
bosons are not visible in observables. For a color
superconductor, it can be expressed as the following: the quark--quark EM
interaction contains an additional term due to the exchange of a
Goldstone boson which is absent in the symmetric phase. For example,
the lowest quark-quark scattering T-matrix is of the following
form
\begin{eqnarray}
      T = (iJ_{core})_\mu G_T^{\mu\nu} (iJ'_{core})_\nu +
      g^2(iJ_{spread}) G_G (iJ'_{spread})
\label{T1}
\end{eqnarray}
where $G_T^{\mu\nu}$ is the full propagator of a photon, $g\sim O(1)$
is the coupling constant between the Goldstone boson and the quarks
and $G_G$ is the propagator of the Goldstone boson. Then Ward
identities that specifies the right hand side of Eq. \ref{test2} and
others can be used to recast Eq. \ref{T1} into the following form
\begin{equation}
     T = (iJ_{core})_\mu \left (G_T^{\mu\nu} + G_L^{\mu\nu} 
         \right )(iJ'_{core})_\nu 
\label{T2}
\end{equation}
with the vacuum
\begin{equation}
    G^{\mu\nu} = {-g^{\mu\nu} + q^\mu q^\nu/m_\gamma^2\over q^2-m_\gamma^2}
\end{equation}
a propagator for a genuine massive vector particle that couples to the
quark with the strength of the electromagnetism $\alpha_{em}\sim
1/137$.  It should be noted that the scattering between two quarks
generated by the Goldstone boson, like pions, is a strong interaction
with strength $g\sim 1$. From this it can be concluded that
$G_T^{\mu\nu}$ also contains a component of strong interaction, which
must cancel the $g^2 G_G$ term. The fact that quark--quark interaction
remains small in an EM interaction depends on a delicate cancellation
of strong interaction effects that are responsible for the symmetry
breaking.

 Things are not that simple in the case of hadronic color
superconductivity since the realistic order parameters for it does not
break the EM $U(1)$ gauge symmetry directly. It breaks the global
$U(1)$ symmetry corresponding the nucleon number (1/3 of the fermion
number for the up and down quarks), which is itself a component of the
baryon number, under the most natural scenario of color
superconductivity \cite{review}. It is induced by a condensation of
diquarks made up of light up and down quarks. In this case, the Higgs
mechanism of cancelling the effects of the Goldstone boson is
effective only in the subspace of normal hadrons made up of up and
down quarks. Other hadrons and all leptons do scatter with normal
hadrons through exchange of the (or the lack of the) Goldstone bosons
of the $U(1)$ symmetry breaking connected to the nucleon number.

 Let us consider the lowest order (in the EM coupling) lepton--nucleon
scattering interested here, which can be written as
\begin{equation}
     T(l+h\to l'+h') = (ij)_\mu G_T^{\mu\nu} (iJ'_{core})_\nu 
\label{T3}
\end{equation}
with $j_\mu$ the matrix element of the EM current operator and
$J'_\mu$ that of the hadrons (or up/down quarks). The Goldstone boson
does not contribute to the lepton-quark scattering since leptons carry
zero nucleon charge. Therefore, using the relation $G_T^{\mu\nu}=
G^{\mu\nu} - G_L^{\mu\nu}$ we have
\begin{equation}
     T(l+h\to l'+h') = T_{em}+ T_{strong}= (ij)_\mu G^{\mu\nu}(iJ'_{core})_\nu - 
                 (ij)_\mu G_L^{\mu\nu} (iJ'_{core})_\nu
\label{T4}
\end{equation}
where $T_{em}$ is a small EM interaction scattering kernel
corresponding to the first term on the right hand side of the above
equation and the rest, $T_{strong}$, is an induced strong interaction
kernel  due to the lack of the Goldstone boson contributions
between lepton and quarks in the color superconducting phase.

 It implies that there is a significant strong interaction component
in the semi-leptonic DIS that is absent in the normal hadron--hadron
collision due to the Higgs cancellation in the later situations if a
hadron has a ``superconducting component'' or if there is a close by
virtual color superconducting phase for the strong interaction vacuum.

\section{The Froissart Bound in DIS} 

 The Froissart bound requires that the total cross section for a 
hadron--hadron collision at high enough 
energy must satisfies
\begin{equation}
   \sigma^{hh}_{tot} \le \mbox{const}\times \ln^2 s
\label{Fb}
\end{equation}
with $s$ the energy of the hadron involved. It turns out that the
physical hadron--hadron total cross section do indeed satisfies the
bound: in fact it lies just on the bound. In Regge theory, the reggeon
exchanged that is responsible for the behavior is called the soft
pomeron with an intercept of $\alpha_{\cal P} \propto 1.08$.

 The total cross section for a $\gamma^* N$ interaction extracted from the
semi-leptonic DIS appears to grow much faster than the right hand side of
Eq. \ref{Fb}, in fact
\begin{equation}
    \sigma^{\gamma^* N} \sim s^{0.4},
\label{vFb}
\end{equation}
which, in the language of the Regge theory, corresponds to an exchange
of a hard pomeron with a intercept of $\alpha_{{\cal P}'}\approx
1.4$.  What cause such a higher rate of particle production remains much
of a mystery from the unitarity point of view,  albeit perturbative QCD
can explain some of them, some problems remain \cite{Ball}. It is not sure
the spin content of a nucleon can be simultaneously explained \cite{Reya}.

The assumption that there is at least one color superconducting phase
for the strong interaction vacuum state makes such a behavior very
natural. Since the extra final states produced is due to the exchange
of the (lack of) Goldstone boson of the color superconducting phase.

\section{Empirical evidences}

 The consequences of the assumption that a nucleon contains a
superconducting companion which lives in the virtual color
superconducting phase can be subjected to a variety of tests
\cite{review} using the known data. A partial list of them are 1)
violation of GDH sum rule which is still not eliminated by the current
data \cite{gdht} 2) charge separation (Eq. \ref{test1}) manifested in
the difference \cite{DISpap} between the structure function $F_2$
extracted from the charged lepton neutral current DIS and from the
neutrino charged current DIS at small Bjorken x ($x<0.1$) 3) the
appearance \cite{DISpap} of a photon-like vector pomeron coupled to a
non-conserved current (Eq. \ref{test2}) in the meson production
processes of a NN collision 4) the violation of the Froissart bound
(Eq. \ref{vFb}) discussed above and the more detailed predictions
concerning the different energy dependences of the electro-production
of the $\rho$, $\phi$ and $J/\psi$ mesons under the hard pomeron
hypothesis \cite{Mprod}. They are all in qualitative agreement with the
data.  Under the hard pomeron hypothesis, the small x behavior of the
polarized structure functions $g_1$ and $g_2$ of a nucleon have the
following small but quite singular component at small x before the
Bjorken limit
\begin{eqnarray}
     g_1(x) &\sim & {1\over x} F_2(x) \sim x^{-1.4}, \hspace{1.5cm}
     g_2(x) \sim {1\over x^2} F_2(x) \sim x^{-2.4}
\end{eqnarray}
which comes from the superconducting component of a nucleon mentioned
above. The first equation above is consistent \cite{DISpap} with the
current SMC data that has its smallest $x\sim 10^{-4}$.  Such a
behavior is hard to explain in a perturbative QCD scheme. 

Albeit many evidences of quite different origin are correlated to
point to the possibility that there is at least one close by virtual
color superconducting phase for the strong interaction vacuum, more
work are needed to firmly establish it \cite{review}.

\section*{Acknowledgements}

   This work is supported by the National Natural Science Foundation
of China. I would like to thank Prof. A. W. Thomas and the CSSM of
Univeristy of Adelaide for hospitality during which this write up is
done.

\end{document}